\documentclass[rmp,aps,amsfonts,amsmath,amssymb,nofootinbib,twocolumn,superscriptaddress,numerical]{revtex4-2}
\setcitestyle{numbers,square,sort&compress}
\usepackage{graphicx}
\usepackage{bm}
\usepackage{hyperref}

\hypersetup{
    colorlinks,
    linkcolor={red!50!black},
    citecolor={blue!50!black},
    urlcolor={blue!80!black}
}

\usepackage{float}
\usepackage{xcolor}

\graphicspath{{figures/}}
\begin{document}

\title{Development of Biphoton Entangled Light Spectroscopy (BELS) using Bell pairs }

\author{V.V. Desai}
\affiliation{William H. Miller III Department of Physics and Astronomy, The Johns Hopkins University, Baltimore, Maryland 21218, USA}

\author{N.P. Armitage}
 \email{npa@jhu.edu}
\affiliation{William H. Miller III Department of Physics and Astronomy, The Johns Hopkins University, Baltimore, Maryland 21218, USA}

\date{\today}

\maketitle

\textbf{We introduce Biphoton Entanglement Light Spectroscopy (BELS), a quantum spectroscopic technique that employs polarization-entangled Bell pairs and two-photon interference to probe material properties. In BELS, the measured signal arises not from single-photon intensities but from changes in the joint polarization and path correlations of biphoton Bell pairs transmitted through or scattered by a sample and analyzed via cross-channel coincidences. A key concept of BELS is the explicit mapping between Jones matrix operations and transformations within the Bell-state manifold. Optical elements that are equivalent under classical polarization optics can produce qualitatively distinct signatures in the coincidence landscape when interrogated with entangled photons. We demonstrate that linear birefringence and Faraday rotation generate orthogonal admixtures of Bell states, yielding experimentally distinguishable coincidence channels within a single measurement. We measure birefringence in an anisotropic dielectric and Faraday rotation in Tb$_3$Ga$_5$O$_{12}$. By mapping the changes to the photonic entanglement, BELS establishes a new framework for future entanglement enhanced spectroscopy—a potentially powerful approach in characterizing quantum materials, nanophotonic devices, and light–matter interactions perhaps eventually at a fundamentally quantum level.}


\bigskip
 
\bigskip

\section{Introduction}

Quantum technologies including computing, communication, sensing, and imaging harness entanglement, superposition, and nonlocal correlations to surpass classical limits.~\cite{preskill2018quantum,gisin2002quantum,giovannetti2004quantum, brida2010experimental}.  They offer advantages in precision, sensitivity, efficiency, and information security~\cite{rahim2023quantum}. 

It has also been increasingly recognized that there are materials and phenomena at the forefront of modern condensed matter physics such as spin liquids, quantum criticality, and thermalization in many-body systems~\cite{balents2010spin,grover2013entanglement} that are believed to have properties intimately related to quantum mechanical entanglement, but there are few probes to measure it.  Optical spectroscopy is in general an important tool to characterize materials.  However even for materials where the quantum properties are paramount, the characterization has been done with classical light~\cite{basov2011electrodynamics}.  In this respect it is important to develop methods to measure material properties with entangled photons, with an eye towards eventually measuring intrinsically {\it quantum} properties of materials with entangled photons.  Moreover, one may hope that even the measurement of classical properties with quantum entangled photons may afford advantages similar to that in other quantum technologies.

In this work we present a new scheme of Bell pair Biphoton Entangled Light Spectroscopy (BELS) to measure the optical response of materials.  The method uses polarized entangled Bell pairs inputted into a Hong-Ou-Mandel (HOM) interferometer and measures sample properties placed in one beam path.  The measurement signal arises not from single- or multi-photon intensity as in most spectroscopies, but from changes in the two photon interference properties of biphoton Bell pairs transmitted through or scattered by a sample and measured by cross coincidences between path and polarization channels.  We demonstrate that linear birefringence and Faraday rotation produce orthogonal admixtures of Bell states, leading to experimentally distinguishable coincidence channels of these two phenomena within a single measurement configuration.  We give an explicit demonstration of the use of this technique to characterize Faraday rotation on the material Tb$_3$Ga$_5$O$_{12}$.

The Hong-Ou-Mandel (HOM) interferometer is a classic experiment~\cite{hong1987measurement} in quantum optics, which has been used to demonstrate  quantum mechanical entanglement~\cite{ou1988violation,shih1988new} quantum teleportation~\cite{bouwmeester1997experimental}, implementation of CNOT gates~\cite{o2003demonstration}, and quantum computing operations~\cite{kok2007linear}.  Physically it is a variant on an optical beamsplitter.  When two classical particles enter the two input ports there are 4 possible outcomes.   Both can be transmitted, both can be reflected, or one can be reflected and the other transmitted.  In contrast, when two indistinguishable photons enter the two input ports, quantum mechanics requires that both photons exit from the same output port\footnote{This arises from the fact that the conversion matrix for a lossless 50/50 beamsplitter must be unitary to conserve probability~\cite{zeilinger1981general,holbrow2002photon}.  The matrix can be expressed as $\frac{1}{\sqrt{2} } \left[\begin{array}{cc} r & t' \\ t & r'\end{array}\right] $ with the coefficients expressed as complex exponentials e.g. $r = |r| e^{i \delta_r}$.  The condition for unitarity is that the Hermitian adjoint equals the inverse.  If we set the determinant of the matrix to 1 (equivalent to choosing a global phase factor of the conversion matrix to zero), we have the condition that $t = -t'^*$ and $r=r'^*$.   Dividing one expression by the other one finds $\delta_t  - \delta_r + \delta_{t'} - \delta_{r'} = \pi$.   For a symmetric beamsplitter $r=r'$ and $t=t'$, which requires $\delta_t  - \delta_r = \delta_{t'} - \delta_{r'} = \pi/2$.  For asymmetric beams splitters there are other choices. Eq. \ref{beamsplitter} is one possibility. } e.g. the probability amplitude for one photon to be transmitted and the other to be reflected destructively interferes with the probability amplitude for the reverse case. This non-classical interference of few-photon wavefunctions is observed by measuring coincidences at detectors placed at the two output ports.  HOM interference is demonstrated by varying the time delay between the arrival of the two photons at the beamsplitter.  The probability that photons arrive in coincidence at separate detectors at the two output ports goes to zero when the relative delay is zero e.g. they arrive at the beamsplitter at the same time and are indistinguishable, which causes the celebrated HOM ``dip" in the coincidence rate (see Fig.~\ref{HOM3mm}).

\begin{figure}[t]
	\includegraphics[width=0.45\textwidth]{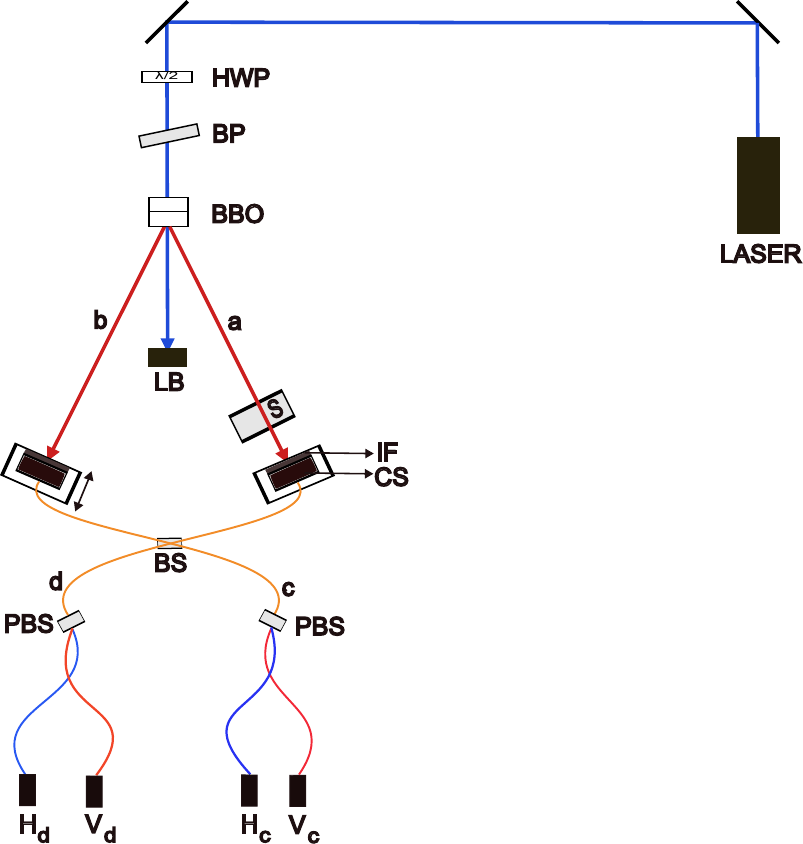}
	\caption{\textbf{Modified Hong–Ou–Mandel Interferometer for Biphoton Entanglement Light Spectroscopy (BELS).}
 The scheme incorporates entangled Bell pairs and polarization detection.  HWP - half wave plate, BP - birefringent plate, BBO -  $\beta$-Barium Borate, LB - Laser beam Block, S- sample, IF - Interference filter, CS - Collimation system, BS- 50/50 beamsplitter, PBS - polarizing beam splitter.   In arm $b$ a dielectric compensator (fused silica or a HWP) was introduced to match the optical thickness of the sample placed in arm $a$ (not shown).     }
 \label{HOM}
\end{figure}

\begin{figure*}[t]
	\includegraphics[width=0.7\textwidth]{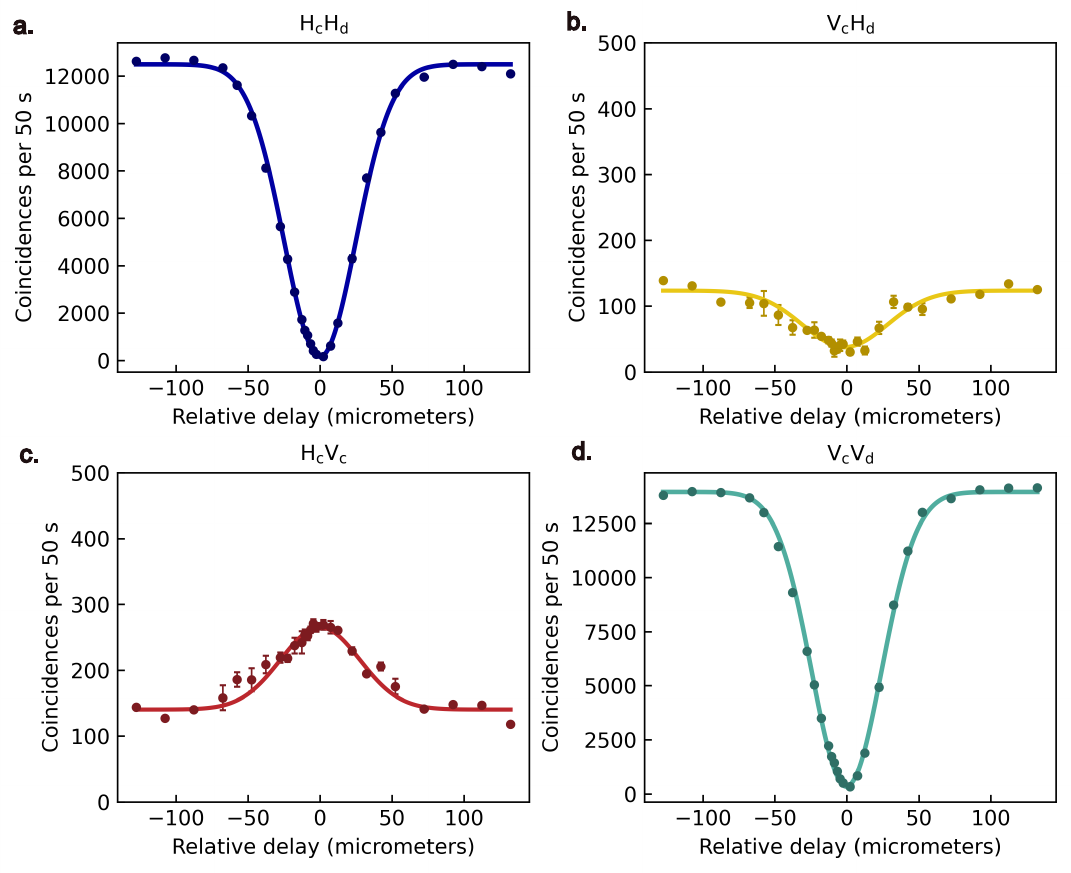}
	\caption{\textbf{Polarization-Resolved HOM Interference for a Nominal $| \Phi^+ \rangle$ State}. Coincidences as a function of relative arm delay for coincidence channels for a nominal $| \Phi^+ \rangle$ input state for a pair of 3mm orthogonally attached BBO crystals for Type I SPDC.  (a) $\mathrm{H}_{\mathrm{c}}$:$\mathrm{H}_{\mathrm{d}}.$   (b) $\mathrm{V}_{\mathrm{c}}$:$\mathrm{H}_{\mathrm{d}}.$   (c)  $\mathrm{H}_{\mathrm{c}}$:$\mathrm{V}_{\mathrm{c}}.$   (d) $\mathrm{V}_{\mathrm{c}}$:$\mathrm{V}_{\mathrm{d}}$.   Note that vastly different scales of (a) and (d) as compared to (b) and (c)   Fits are Gaussian plus a constant.   }
 \label{HOM3mm}
 \end{figure*}

Now consider a modified version of a HOM interferometer shown schematically in Fig. \ref{HOM}.   It differs from the usual scheme in that instead of sending in two photons along paths $a$ and $b$ in a separable wavefunction, one sends in two polarization entangled photons in one of the 4 Bell states along arms $a$ and $b$ and then detecting coincidences in the polarizations of photons that end up at polarization resolved detectors in arms $c$ and $d$.  The possible states are spanned by a 10 dimensional Hilbert space, but we will restrict the input states to the Bell states which span a more limited 4 dimensional space.  The four Bell states are

\begin{align}
| \Psi^\pm\rangle &= \frac{1}{\sqrt{2} } (|H\rangle_a |V\rangle_b \pm |V\rangle_a |H\rangle_b ), \\
| \Phi^\pm\rangle &= \frac{1}{\sqrt{2} } (|H\rangle_a |H\rangle_b \pm |V\rangle_a |V\rangle_b ).
\end{align}

As discussed above, the properties of a unitary beamsplitter are such that the relative phases of the output modes are constrained.  For an asymmetric dielectric beamsplitter the output modes may be related to the input modes as



\begin{equation}
\frac{1}{\sqrt{2} } \left[\begin{array}{cc} 1 & 1 \\ - 1 & 1\end{array}\right] 
\left[\begin{array}{c} a \\ b \end{array}\right]  = \left[\begin{array}{c}c \\ d \end{array}\right].
\label{beamsplitter}
\end{equation}

For an ideal beamsplitter this expression applies to both $|H \rangle$ and $|V\rangle$ polarized light in input channels $a$ and $b$.   With one of the 4 Bell states as input, there are four output states from the beam splitter:

\begin{align}
| \Psi^+\rangle  \xrightarrow{BS} \frac{1}{\sqrt2} \Big (|HV\rangle_c | 0 \rangle_d - | 0 \rangle_c | HV \rangle_d \Big ),
\label{BSpsi+}
\end{align}

\begin{align}
| \Psi^-\rangle \xrightarrow{BS} - \frac{1}{\sqrt2} \Big (|H\rangle_c | V \rangle_d - | V \rangle_c | H \rangle_d \Big ), 
\label{BSpsi-}
\end{align}

\begin{align}
| \Phi^\pm\rangle  \xrightarrow{BS}  \frac{1}{2\sqrt2} \Big (|HH\rangle_c | 0 \rangle_d \pm | V V \rangle_c | 0 \rangle_d-  &  \nonumber \\   | 0 \rangle_c | HH \rangle_d \mp  &  | 0 \rangle_c | VV \rangle_d \Big ).
\label{BSphi}
\end{align}
The output states are detected by measuring coincidences between the detectors $\mathrm{H}_{\mathrm{c}}$, $\mathrm{V}_{\mathrm{c}}$, $\mathrm{H}_{\mathrm{d}}$ and  $\mathrm{V}_{\mathrm{d}}$~\footnote{This is slightly unconventional notation where $|HH\rangle_c$ means two H photons in the $c$ channel and $|HV\rangle_c$ means an H and V photon in the $c$ channel.}. A component of $| \Psi^+\rangle$ manifests by giving coincidences in $\mathrm{H}_{\mathrm{c}}$:$\mathrm{V}_{\mathrm{c}}$ and $\mathrm{H}_{\mathrm{d}}$:$\mathrm{V}_{\mathrm{d}}$. A component of  $| \Psi^-\rangle$ gives coincidences in $\mathrm{H}_{\mathrm{c}}$:$\mathrm{V}_{\mathrm{d}}$  and $\mathrm{V}_{\mathrm{c}}$:$\mathrm{H}_{\mathrm{d}}$.  This is an anti-bunching effect between channels that arises because the $| \Psi^-\rangle$ state, while being symmetric overall, is anti-symmetric with respect to paths and polarization separately.  Notably,  the $| \Phi ^\pm\rangle$ states show coincidences similar to the usual HOM effect and gives no cross channel coincidences.   The analysis bears some resemblance to that implemented in the double HOM setup applied to momentum entangled Bell states in Ref.~\cite{michler1996interferometric}.

We now consider the effect of placing a sample in arm $a$ of the interferometer.   The generic effect of a sample on a coherent beam is given by a ``Jones" matrix~\cite{armitage2014constraints}.  The most general Jones matrix can be written as a superposition of orthogonal matrices e.g.

\begin{align}
\hat{T} =  \left[\begin{array}{cc} A & B \\ C & D\end{array}\right] = & \nonumber \\  \alpha \left[\begin{array}{cc} 1 & 0 \\ 0 & 1\end{array}\right] +  \beta  \left[\begin{array}{cc} 1 & 0 \\ 0 & -1\end{array}\right] +    \kappa    \left[\begin{array}{cc} 0 & 1 \\ 1 & 0 \end{array}\right]  & +  \delta \left[\begin{array}{cc} 0 & 1 \\ -1 &  0 \end{array}\right].
\label{Jones}
\end{align}

Often materials are governed by particular symmetries which set some of the coefficients of Eq. \ref{Jones} to zero~\cite{armitage2014constraints}.   To understand the effect of the sample Jones matrix $\hat{T}$ on the Bell states, we can start with understanding the effect of each term in Eq. \ref{Jones}.  The action of the unit matrix is trivial.  The others can be easily calculated using the Jones matrix formalism where it acts on a photon in the $a$ path\footnote{To give an explicit example from Eqs. \ref{quarter} we have
\begin{gather}
 \left[\begin{array}{cc} 0 & 1 \\ -1 &  0 \end{array}\right]_a |\Phi^\pm \rangle =   \frac{1}{\sqrt{2}} 
  \left[\begin{array}{cc} 0 & 1 \\ -1 &  0 \end{array}\right]_a   \Big ( |H\rangle_a |H\rangle_b \pm  |V\rangle_a |V\rangle_b \Big  )  = \nonumber \\ \frac{1}{\sqrt{2}} 
 \left[\begin{array}{cc} 0 & 1 \\ -1 &  0 \end{array}\right]_a \Bigg[ \left[\begin{array}{c} 1 \\ 0 \end{array}\right]_a \left[\begin{array}{c} 1 \\ 0 \end{array}\right]_b \pm \left[\begin{array}{c} 0 \\ 1 \end{array}\right]_a \left[\begin{array}{c} 0 \\ 1 \end{array}\right]_b \Bigg ] = \nonumber \\ 
 \frac{1}{\sqrt{2}} \Bigg [ - \left[\begin{array}{c} 0 \\ 1 \end{array}\right]_a \left[\begin{array}{c} 1 \\ 0 \end{array}\right]_b \pm \left[\begin{array}{c} 1 \\ 0 \end{array}\right]_a \left[\begin{array}{c} 0 \\ 1 \end{array}\right] \Bigg ]_b  = \nonumber  \\ \frac{\pm 1 }{\sqrt{2}} 
(|H\rangle_a |V\rangle_b \mp |V\rangle_a |H\rangle_b ) 
 = \pm |\Psi^\mp \rangle.
\end{gather}}.   One finds

\begin{align}
 \left[\begin{array}{cc} 1 & 0 \\ 0 &  -1 \end{array}\right]_a |\Psi^\pm \rangle & =| \Psi^\mp \rangle , \; \; \;
  \left[\begin{array}{cc} 1 & 0 \\ 0 &  -1 \end{array}\right]_a |\Phi \rangle^\pm = |\Phi^\mp\rangle,
\end{align}

\begin{align}
 \left[\begin{array}{cc} 0 & 1 \\ 1 &  0 \end{array}\right]_a |\Psi ^\pm\rangle = \pm | \Phi ^\pm\rangle  , \; \; \;   \left[\begin{array}{cc} 0 & 1 \\ 1 &  0 \end{array}\right]_a |\Phi ^\pm\rangle =  \pm|\Psi ^\pm\rangle ,
 \label{half}
\end{align}

\begin{align}
 \left[\begin{array}{cc} 0 & 1 \\ -1 &  0 \end{array}\right]_a |\Psi^\pm\rangle = \pm |\Phi^\mp\rangle  , \; \; \;
  \left[\begin{array}{cc} 0 & 1 \\ -1 &  0 \end{array}\right]_a | \Phi ^\pm\rangle =  \pm|\Psi ^\mp\rangle .
   \label{quarter}
\end{align}

 \begin{figure}[t]
	\includegraphics[width=0.45\textwidth]{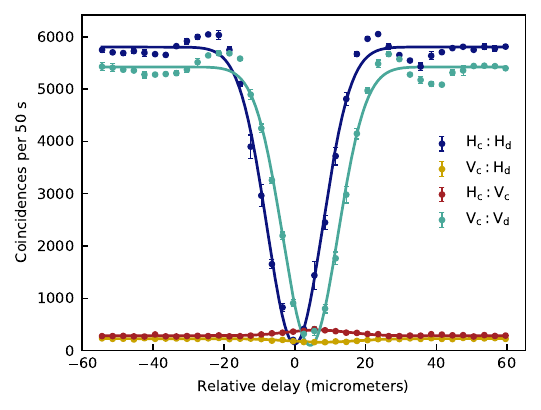}
	\caption{\textbf{Biphoton Coherence as a function of relative delay with Coincidence Landscapes.} Coincidences for channels $\mathrm{H}_{\mathrm{c}}$:$\mathrm{H}_{\mathrm{d}}$ (Minima; 124, Plateau; 5806),   $\mathrm{V}_{\mathrm{c}}$:$\mathrm{H}_{\mathrm{d}}$ (Minima; 159, Plateau; 228) ,   $\mathrm{H}_{\mathrm{c}}$:$\mathrm{V}_{\mathrm{c}}$ (Maxima; 393, Plateau; 282),   $\mathrm{V}_{\mathrm{c}}$:$\mathrm{V}_{\mathrm{d}}$ (Minima; 98,  Plateau; 5425 ) for Type I SPDC for a pair of 0.5 mm BBO crystals and a 30 nm interference filter. Fits are Gaussian plus a constant.    }
 \label{HOM05mmoverlay}
 \end{figure}

Notably ``perfect" optical elements like half- and quarter wave plates (represented by the third and fourth matrices of the decomposition in Eq. \ref{Jones} and their effects by Eqs. \ref{half} and \ref{quarter}) will cause complete conversion from one Bell state to another.   However most materials will be represented by a sum of terms in Eq. \ref{Jones} with typically the the unit matrix dominating.  If one starts with a nominally pure Bell state and introduces a sample into one arm, it will have the effect of mixing into the photonic wavefunction one of the other Bell states.  As an example, let us consider a sample that introduces a pure Faraday rotation.   Its transmission matrix projected on the Jones matrix decomposition is

\begin{align}
 \left[\begin{array}{cc} \mathrm{cos} \; \theta & \mathrm{sin}  \; \theta \\ - \mathrm{sin}  \; \theta & \mathrm{cos}  \; \theta \end{array}\right] =  \mathrm{cos}  \; \theta \left[\begin{array}{cc} 1 & 0 \\ 0 & 1\end{array}\right] +   \mathrm{sin} \; \theta \left[\begin{array}{cc} 0 & 1 \\ -1 &  0 \end{array}\right].
\label{Rotation}
\end{align}
e.g. $\alpha = \mathrm{cos}  \; \theta $ and $\delta = \mathrm{sin}  \; \theta $.  If one starts with the generated Bell state $|\Phi^+ \rangle$, then after the sample one has a state 

\begin{align}
|\Upsilon \rangle = \alpha |\Phi^+ \rangle + \delta |\Psi^- \rangle
\label{Wavefunction}
\end{align}
The effect of the sample is to mix in $| \Psi^- \rangle $.  As discussed above, the state  $|\Phi^+\rangle $ generates no cross-channel coincidences after propagating through the HOM interferometer.  However $|\Psi^- \rangle $ gives coincidences in $\mathrm{H}_{\mathrm{c}}$:$\mathrm{V}_{\mathrm{d}}$ and $\mathrm{V}_{\mathrm{c}}$:$\mathrm{H}_{\mathrm{d}}$.  Consider a situation where the Faraday angle $\theta $ grows continuously say as a function of temperature due to a magnetic phase transition.   The parameter $\delta$ in Eq. \ref{Jones} will be proportional to the order parameter and hence the coincidence signal will grow below the magnetic transition as $\delta^2$.  One of the remarkable things about this scheme is to note that it can detect the difference between Faraday rotations and linear birefringence in a single measurement, as birefringent element will mix in $ | \Psi^+ \rangle$ and give a different set of coincidences $(\mathrm{H}_{\mathrm{c}}$:$\mathrm{V}_{\mathrm{c}}$ and $\mathrm{H}_{\mathrm{d}}$:$\mathrm{V}_{\mathrm{d}}$).   Such a distinction usually involves multiple measurements with conventional optics, whereas here it only involves a single measurement.   This is reminiscent of the situation in quantum computation where many channels of information can be processed in parallel.

\section{Results}

We have implemented the scheme shown in Fig.~\ref{HOM} in a physical spectrometer.  We used an 405 nm laser InGaN diode laser incident on a $\beta$-Barium Borate (BBO) crystal to create two polarization entangled 810 nm photons via Type I spontaneous parametric down conversion (SPDC).  Important for the below discussion is that the fiberport coupler in arm $b$, which couples the free space beam into a polarization maintaining single mode fiber can be moved along the beam axis allowing one to tune the relative delay of the two paths which allow us to make the photons indistinguishable or not.   A more detailed overview of the apparatus is given in {\bf METHODS} below.

In Figs. \ref{HOM3mm}a - \ref{HOM3mm}d we show coincidences for  four separate channels for a pair of 3 mm orthogonally attached BBO crystals for Type I SPDC and a 20 nm interference filter.  For a pure $ |\Phi^+ \rangle$ state, the relative probability of detecting
$\mathrm{H}_{\mathrm{c}}$:$\mathrm{H}_{\mathrm{d}}$ or $\mathrm{V}_{\mathrm{c}}$:$\mathrm{V}_{\mathrm{d}}$ coincidences is $0.25$ when the photons can be distinguished e.g. when they arrive with a finite relative delay, which we see as the plateau at larger relative delay on Figures \ref{HOM3mm}a and \ref{HOM3mm}d.  As the relative delay approaches zero, this probability goes towards zero consistent with Eq. \ref{BSphi}.  Consequently, the coincidence rate in the $\mathrm{H}_{\mathrm{c}}$:$\mathrm{H}_{\mathrm{d}}$ and $\mathrm{V}_{\mathrm{c}}$:$\mathrm{V}_{\mathrm{d}}$ channels exhibits a Hong-Ou-Mandel dip as a function of relative delay.  Simultaneously, very small but not zero coincidences are observed in the $\mathrm{V}_{\mathrm{c}}$:$\mathrm{H}_{\mathrm{d}}$ and $\mathrm{H}_{\mathrm{c}}$:$\mathrm{V}_{\mathrm{c}}$ channels.   These largely arise from small amounts of a $|\Psi^+ \rangle$ admixture in the original state, which will cause bunching in the $\mathrm{H}_{\mathrm{c}}$:$\mathrm{V}_{\mathrm{c}}$ channel and anti-bunching in the $\mathrm{V}_{\mathrm{c}}$:$\mathrm{H}_{\mathrm{d}}$ channel.  We do not find a measurable $ |\Psi^- \rangle$ contribution.  From this plot one cannot distinguish between $| \Phi^+ \rangle$ and  $ |\Phi^- \rangle$ states, but this is done below.  The finite residual coincidences at zero relative delay arise from incomplete photon indistinguishability due to the spectral bandwidth, residual birefringent walk-off, and non-ideal spatial mode overlap.

The performance of the spectrometer can also be benchmarked through this polarization-resolved HOM interference from the measured coincidence suppression in the $\mathrm{H}_{\mathrm{c}}$:$\mathrm{H}_{\mathrm{d}}$  and  $\mathrm{V}_{\mathrm{c}}$:$\mathrm{V}_{\mathrm{d}}$ channels.  We extract a HOM dip visibility by taking the ratio of the amplitude of the Gaussian fit and the value of coincidences at the plateau of $\mathcal{V}_{\mathrm{HOM}} = 0.981 \pm 0.001 $ demonstrating high two-photon indistinguishability. 
The corresponding full width at half minimum (FWHM) of the HOM dip yields a biphoton coherence length $\ell_c = 59 \pm 1\ \mu\mathrm{m}$ (coherence time of $ \tau_c = \frac{\ell_c}{c} = 197 \pm 3\ \mathrm{fs} $) consistent with the $\Delta \lambda = 10$ nm bandwidth imposed by the interference filters used in the experiment ($\ell_c = \frac{\lambda^2}{2 \pi \Delta \lambda}$).

In Fig. \ref{HOM05mmoverlay} we show the coincidences for channels $\mathrm{H}_{\mathrm{c}}$:$\mathrm{H}_{\mathrm{d}}, \mathrm{V}_{\mathrm{c}}$:$\mathrm{H}_{\mathrm{d}}, \mathrm{H}_{\mathrm{c}}$:$\mathrm{V}_{\mathrm{c}},$ and $\mathrm{V}_{\mathrm{c}}$:$\mathrm{V}_{\mathrm{d}}$ now generated by Type I SPDC for a different pair of 0.5 mm BBO crystals and a 30 nm interference filter, which was chosen to show different possibilities for operating configurations.  One sees similar coincidences as in Fig.~\ref{HOM3mm}, in that one sees that the $\mathrm{H}_{\mathrm{c}}$:$\mathrm{H}_{\mathrm{d}}$ and $ \mathrm{H}_{\mathrm{c}}$:$\mathrm{V}_{\mathrm{c}}$ coincidences go to near zero when the relative delay is near zero.   This again indicates that the photon wavefunction is primarily one of the $\Phi$ states.  Again there is a very small but not zero admixture of $| \Psi^+ \rangle$ that appears in the comparison of $\mathrm{V}_{\mathrm{c}}$:$\mathrm{H}_{\mathrm{d}}$ and $\mathrm{H}_{\mathrm{c}}$:$\mathrm{V}_{\mathrm{c}}$ coincidence channels as   $| \Psi^+ \rangle$  causes bunching in the $\mathrm{H}_{\mathrm{c}}$:$\mathrm{V}_{\mathrm{c}}$ channel and anti-bunching in the $\mathrm{V}_{\mathrm{c}}$:$\mathrm{H}_{\mathrm{d}}$ channel.  Again, we do not find a measurable $|\Psi^-\rangle$.  Note that curves $\mathrm{H}_{\mathrm{c}}$:$\mathrm{H}_{\mathrm{d}}$ and $\mathrm{V}_{\mathrm{c}}$:$\mathrm{H}_{\mathrm{d}}$ are slightly offset from each other due to the difference in the lengths of the fast and slow axes taken by the H and V polarized photons.  It differs from Fig.~\ref{HOM3mm} because of change in thickness of the crystal and the birefringent compensation being matched better due to the same.  Also note the regions of interference fringes outside the dip that arise from the sharp spectral truncation imposed by the band-pass filters, which generates sinc-like temporal sidelobes through the Fourier relationship between spectral filtering and temporal interference, analogous to the Gibbs phenomenon~\cite{rarity2024not}.

 \begin{figure}[t]
	\includegraphics[width=0.45\textwidth]{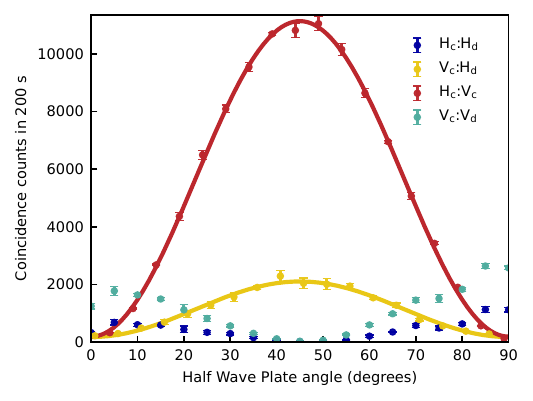}
	\caption{\textbf{Birefringence-Induced Transformations within the Bell-State Manifold.} Scan of 4 coincidences at the zero relative delay position as a function of rotation of a HWP placed in the optical path $a$. Solid lines are fits to the form $\sin^{2}(2\theta)$ with an offset.  $\mathrm{H}_{\mathrm{c}}$:$\mathrm{H}_{\mathrm{d}}$ (0$^\circ$; 783, 45$^\circ$; 34),   $\mathrm{V}_{\mathrm{c}}$:$\mathrm{H}_{\mathrm{d}}$ (0$^\circ$; 175, 45$^\circ$; 2103) ,   $\mathrm{H}_{\mathrm{c}}$:$\mathrm{V}_{\mathrm{c}}$ (0$^\circ$; 191, 45$^\circ$; 11129),   $\mathrm{V}_{\mathrm{c}}$:$\mathrm{V}_{\mathrm{d}}$ (0$^\circ$; 2058,  45$^\circ$; 48 )
    }
 \label{HWP}
 \end{figure}

In Fig.~\ref{HWP}  we show a scan of the same 4 $\mathrm{H}_{\mathrm{c}}$:$\mathrm{H}_{\mathrm{d}}, \mathrm{V}_{\mathrm{c}}$:$\mathrm{H}_{\mathrm{d}}, \mathrm{H}_{\mathrm{c}}$:$\mathrm{V}_{\mathrm{c}},$ and $\mathrm{V}_{\mathrm{c}}$:$\mathrm{V}_{\mathrm{d}}$ coincidences shown in Fig.~\ref{HOM05mmoverlay} at zero nominal delay, but now as a function of rotation of a HWP placed in optical path $a$ in the sample position. Here we use again the pair of 0.5 mm BBO crystals with 30 nm interference filters to generate primarily a $|\Phi^+ \rangle$ state. At a half-wave plate (HWP) angle of $0^\circ$ and zero nominal delay, coincidence rates $\mathrm{V}_{\mathrm{c}}$:$\mathrm{H}_{\mathrm{d}}$ and $\mathrm{H}_{\mathrm{c}}$:$\mathrm{V}_{\mathrm{c}}$ are very small, which is consistent with only small admixtures of $\Psi$ states.   We see appreciable $\mathrm{H}_{\mathrm{c}}$:$\mathrm{H}_{\mathrm{d}}$   and $\mathrm{V}_{\mathrm{c}}$:$\mathrm{V}_{\mathrm{d}}$
coincidences even at zero nominal delay due to the time offsets for $|HH\rangle$ and $|VV\rangle$ production mentioned above for Fig.~\ref{HOM05mmoverlay}.  When the HWP is rotated to $45^\circ$, the part of the photon wavefunction in the $a$ path swaps its character from $H$ to $V$ and vice versa.  A primarily $|\Phi^+\rangle$ state becomes $|\Psi^+\rangle$ and will generate coincidences in the $\mathrm{H}_{\mathrm{c}}$:$\mathrm{V}_{\mathrm{c}}$  channel as represented in Eq. \ref{BSpsi+}.  The peak in $\mathrm{V}_{\mathrm{c}}$:$\mathrm{H}_{\mathrm{d}}$  coincidences arises from a small admixture of $|\Phi^-\rangle$ that was converted into $|\Psi^-\rangle$ when the HWP was rotated to 45$^\circ$.   We also note that a HWP is a birefringent optical element as mentioned earlier and therefore the observed evolution of the coincidence probabilities as a function of the HWP angle constitutes a direct measurement of birefringence-induced ``complete" polarization transformations in the two-photon state.   And again as noted above we can measure the difference between Faraday rotation and birefringence in a single measurement.

From the above data we can make estimates of the purity of a Bell state wavefunction $|\Upsilon \rangle$ created by the SPDC.  $|\Upsilon \rangle$ in the present case was optimized to be largely $|\Phi^+\rangle$, but there could be small contributions from $|\Phi^-\rangle$, $|\Psi^+\rangle$, and $|\Phi^-\rangle$.  The quadrature sum of the coefficients of $|\Psi^+\rangle$ and $|\Psi^-\rangle$ wavefunctions is equal to sum of the plateau values of $\mathrm{H}_{\mathrm{c}}$:$\mathrm{H}_{\mathrm{d}}$ and $\mathrm{V}_{\mathrm{c}}$:$\mathrm{V}_{\mathrm{d}}$  ratioed to plateau values plus the difference in the extrema of $\mathrm{V}_{\mathrm{c}}$:$\mathrm{H}_{\mathrm{d}}$ and $\mathrm{H}_{\mathrm{c}}$:$\mathrm{V}_{\mathrm{c}}$.  This assumes that the residual counts at the zero relative delay are small enough to be neglected.  The square of the coefficient of the admixture of $|\Psi^+ \rangle$ state can be read off from the difference of the extrema on Fig.~\ref{HOM05mmoverlay} ratioed to the denominator.  Finally, the ratio of the squared coefficients of the $|\Phi^+\rangle$ and $|\Phi^-\rangle$ states (with the HWP set to 0$^\circ$) can be read off from the ratio of the $\mathrm{V}_{\mathrm{c}}$:$\mathrm{H}_{\mathrm{d}}$ and $\mathrm{H}_{\mathrm{c}}$:$\mathrm{V}_{\mathrm{c}}$ coefficents with the HWP set to 45$^\circ$.  This gives $ |\langle \Phi^+|\Upsilon\rangle |^2 \approx$ 0.82,  $ |\langle \Phi^-|\Upsilon\rangle |^2 \approx$ 0.15, and  $ |\langle \Psi^+|\Upsilon\rangle |^2 \approx$ 0.03.   We measure no appreciable $| \Psi^- \rangle $.

\begin{figure}[t]
		\includegraphics[width=0.45
	\textwidth]{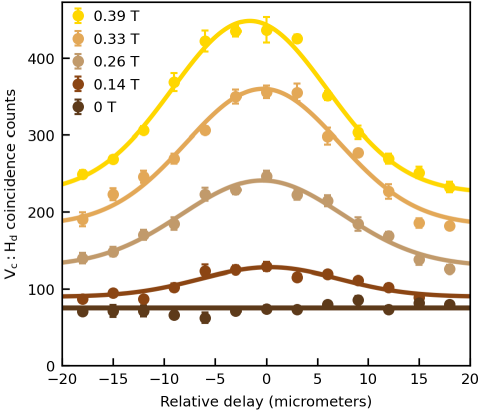}	
	\caption{\textbf{Magnetic-Field-Induced Admixture of $|\Psi^{-}\rangle$ in Tb$_3$Ga$_5$O$_{12}$.} $\mathrm{V}_{\mathrm{c}}$:$\mathrm{H}_{\mathrm{d}}$  coincidence counts as a function of relative delay for different applied magnetic fields for a 10 mm thick TGG sample placed in beam path~$a$.  Solid line are fits to Gaussian with an offset.   We believe the gradual horizontal shift is due to mechanical flexure of the sample stage in magnetic field. }
 \label{InBField1}
 \end{figure}

 \begin{figure}[t]
		\includegraphics[width=0.45
	\textwidth]{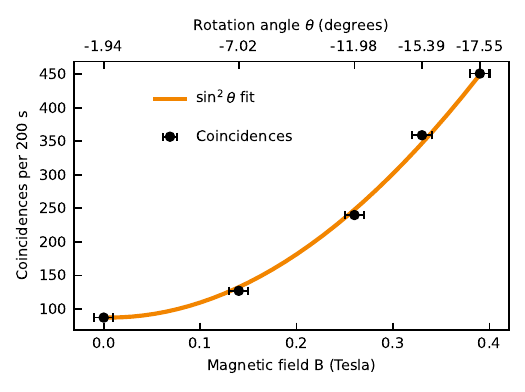}	
	\includegraphics[width=0.45
	\textwidth]{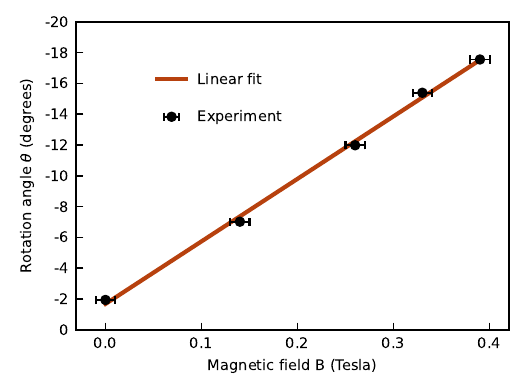}
	\caption{\textbf{Extraction of Faraday Rotation and Verdet Constant from Entanglement-Resolved Coincidences.} (a) $\mathrm{V}_{\mathrm{c}}$:$\mathrm{H}_{\mathrm{d}}$  coincidence counts at the point of zero relative delay as a function of applied magnetic fields for the 10 mm thick TGG sample placed in beam path~$a$.   The solid line is sin$^2 \theta$ plus a constant. (b)  Rotation angle converted from coincidences counts as a function of magnetic field.  }
 \label{InBField2}
 \end{figure}

We further demonstrate our capability by applying BELS to a magneto-optical material, terbium gallium garnet (Tb$_3$Ga$_5$O$_{12}$, TGG).  TGG is a paramagnetic material that is a Faraday rotator with applied magnetic field and induces a magnetic-field-dependent rotation of the plane of linear polarization.  With a 10 mm thick TGG crystal placed in arm $a$ of the interferometer, a small applied magnetic field induces a Faraday rotation, which mixes the $|\Psi^- \rangle $ state into an initially prepared $|\Phi^+ \rangle$ state.   This allows $\mathrm{V}_{\mathrm{c}}$:$\mathrm{H}_{\mathrm{d}}$  and $\mathrm{H}_{\mathrm{c}}$:$\mathrm{V}_{\mathrm{d}}$ coincidences.  In Figure~\ref{InBField1} we show $\mathrm{V}_{\mathrm{c}}$:$\mathrm{H}_{\mathrm{d}}$  coincidence counts as a function of relative delay for different applied magnetic fields, with a TGG crystal placed in beam path~$a$. With an increasing magnetic filed, a progressively greater admixture of the $|\Psi^{-}\rangle$ component is mixed into the initial $|\Phi^{+}\rangle$ state.  This behavior is reflected in the increasing peak values of the $\mathrm{V}_{\mathrm{c}}$:$\mathrm{H}_{\mathrm{d}}$  coincidences with relative delay, consistent with theoretical expectations described by Eqs. \ref{BSpsi-} and \ref{Rotation}.

Figure \ref{InBField2}(a) shows the coincidence counts at zero relative delay, corresponding to the maxima extracted from Fig.~\ref{InBField1}, plotted as a function of the applied magnetic field.  As illustrated by Eq.~\ref{Wavefunction} the effect of the TGG is to mix in $ \delta | \Psi^-\rangle$ into the photon wavefunction $ |\Upsilon\rangle$, which is primarily $  | \Phi^+\rangle$.  The number of coincidences is expected to go like $\delta^2$ and with $\delta $  sin of the rotation angle $\theta$, data are fit to a $\sin^{2}\theta$ dependence derived for an input $|\Phi^{+}\rangle$ state, analogous to the analysis presented in Fig.~\ref{HWP}. From this fit, the polarization rotation angles associated with the measured coincidence counts are extracted and subsequently plotted as a function of magnetic field in Figure \ref{InBField2}(b).  The slope of this dependence, when normalized by the crystal length, yields the Verdet constant ($ \theta(B) = VLB $, where $L$ is the crystal length) of the terbium gallium garnet (TGG) sample. A linear fit to the measured Faraday rotation angle as a function of magnetic field yields a Verdet constant of $V = -71 \pm 2\  \mathrm{rad\ T^{-1}\,m^{-1}}$, which compares favorably to the reported values of $-73 \pm 3\  \mathrm{rad\ T^{-1}\,m^{-1}}$ for TGG at $810\,\mathrm{nm}$ to within experimental uncertainty~\cite{slezak2015wavelength}. Notably, this Faraday rotation is inferred entirely from two-photon entanglement dynamics rather than classical polarization analysis.

\section{Discussion}
In this work we have introduced and experimentally demonstrated BELS as a new spectroscopic framework in which the measured signal arises from changes in two-photon quantum correlations rather than single-photon intensities. By combining polarization-entangled Bell pairs with a modified HOM interferometer, we directly probe how material responses transform photonic entanglement. This departs from conventional spectroscopy, where properties are inferred from intensity-based observables described within classical optics.

A key concept of BELS is the explicit mapping between Jones matrix operations and transformations within the Bell-state manifold. Optical elements that are equivalent at the level of classical polarization optics can induce qualitatively distinct signatures in the coincidence landscape when probed with entangled photons. In particular, we showed that linear birefringence and Faraday rotation produce orthogonal admixtures of Bell states, leading to experimentally distinguishable coincidence channels within a single measurement configuration. This entanglement-resolved sensitivity allows multiple symmetry channels of the optical response to be interrogated simultaneously, rather than sequentially as in conventional polarimetry.

Beyond this specific demonstration, BELS provides a general framework for entanglement-resolved spectroscopy. Two-photon interference techniques may also allow measurement of four-point correlation functions~\cite{irfan2021quantum}, even in systems where the underlying response is classical. Because the signal encodes how a material preserves, converts, or degrades quantum correlations, modified implementations may be suited to systems where quantum coherence, symmetry, or nonreciprocity play central roles. Potential future applications include the investigation of chiral and non-reciprocal photonic devices, magnetically ordered and topological materials, and hybrid quantum systems where light--matter interactions are inherently entanglement-sensitive.

More broadly, BELS illustrates how concepts from quantum information such as Bell-state transformations and coincidence-based measurements can serve as diagnostic tools in condensed-matter and optical spectroscopy. By shifting the focus from classical observables to the response of entanglement itself, this approach opens a path toward spectroscopic techniques capable of probing material properties at a fundamentally quantum level.

\section{Methods}

The spectrometer uses a 405 nm InGaN diode laser that is incident with on a $\beta$-Barium Borate (BBO) crystal for spontaneous parametric down conversion (SPDC). Two orthogonally pasted BBO crystals are used to generate entangled photons using Type I SPDC.   We used both 3mm and 0.5mm pairs in such a configuration.  Two entangled Bell photons are created and fed into polarization-maintained single mode fiber optic cables using fiberport couplers.  Initial alignment was performed using a multimode fiber and a fixed-focus collimation lens to couple the down-converted photons directly into the detectors.  Final measurements employed polarization-maintaining single-mode fibers, whose core diameter is approximately an order of magnitude smaller, necessitating precise alignment.
Efficient coupling was achieved using a Thorlabs PAF2 fiberport coupler, providing translation of the coupling lens along the $x$, $y$, and $z$ axes and angular control via pitch
and yaw. The collimation assembly was mounted on a micrometer translation stage, allowing selective coupling from diametrically opposite regions of the SPDC emission cone to detect entangled photon pairs.
Narrow 10 nm and 30 nm bandpass interference filters are used to spectrally filter the down-converted photons.  The phase of the free-space Bell pair can be adjusted by a single BBO crystal that is placed in the free space beam path either before the SPDC crystal or in one of the beam paths in a similar manner as shown in Fig.~1(b) of Ref. ~\cite{kwiat1999ultrabright}.

The pump-beam HWP (405 nm) is set to $22.5^\circ$, as required for Type-I SPDC.  To prepare the input state in a pure $|\Phi^{+}\rangle$ or $|\Phi^{-}\rangle$ form, distinguishability between the $|\Psi^{+}\rangle$ and $|\Psi^{-}\rangle$ states is first
established, since coincidence measurements alone cannot directly differentiate between the two $\Phi$ states.  Another HWP(810 nm) is inserted in arm~$a$ and set to $45^\circ$, thereby transforming the input state into either $|\Psi^{+}\rangle$ or $|\Psi^{-}\rangle$. At zero relative delay, a birefringent plate (BP - a 0.5 mm BBO crystal) is adjusted until the $\mathrm{H}_{\mathrm{c}}$:$\mathrm{V}_{\mathrm{c}}$ coincidences are maximized while the
$\mathrm{V}_{\mathrm{c}}$:$\mathrm{H}_{\mathrm{d}}$ coincidences are minimized, indicating the
$|\Psi^{+}\rangle$ state, with the opposite behavior corresponding to $|\Psi^{-}\rangle$.
Ideally, the $\mathrm{H}_{\mathrm{c}}$:$\mathrm{H}_{\mathrm{d}}$ and
$\mathrm{V}_{\mathrm{c}}$:$\mathrm{V}_{\mathrm{d}}$ coincidence channels vanish at this point.  Any residual path-length mismatch between the horizontal and vertical polarization components, arising from the fast and slow axes of the beamsplitter, manifests as finite coincidences in the $\mathrm{H}_{\mathrm{c}}$:$\mathrm{H}_{\mathrm{d}}$ or $\mathrm{V}_{\mathrm{c}}$:$\mathrm{V}_{\mathrm{d}}$ channels. Finally, removal of the HWP in arm~$a$ ensures preparation of a photonic wavefunction that is principally $|\Phi^{+}\rangle$.

For the HOM interferometry, a fiber coupled polarization maintaining 50:50 beam splitter from OZ Optics was used, which then couples the beams to two polarizing beam splitters before reaching four Excelitas' Single Photon Counting Module-AQRH detectors. Coincidence detection was done by using a Terasic DE2-115 Field Programmable Gate Array with a coincidence window of 5 ns.  Out of region of the HOM dip, we recorded on the order of $10^{4}$ entangled-photon coincidence counts per $10^{5}$ photons in each arm when using a pair of $3\,\mathrm{mm}$ BBO crystals for Type I spontaneous parametric down-conversion and pumped with $50\,\mathrm{mW}$ of laser power per 50s.
 
Initial alignment was performed using a multimode fiber with a fixed-focus collimation lens to couple the down-converted light directly into the fiber, which was connected to the single-photon detectors. For the final measurements, polarization-maintaining single-mode fibers were required, for which the fiber core diameter is approximately an order of magnitude smaller than that of multimode fibers. Efficient coupling under these conditions necessitates precise control of the coupling optics along all three translational axes. To achieve this, we used a Thorlabs PAF2-series fiber coupler, which provides control of translation along the $x$, $y$, and $z$ axes, as well as angular adjustment via pitch and yaw, enabling optimized mode matching and maximal coupling efficiency. The entire collimation assembly was mounted on a micrometer translation stage, enabling selective coupling of down-converted light from diametrically opposite regions of the SPDC emission cone in order to detect entangled photon pairs. An M30X motorized linear translational stage that has a bidirectional repeatability of 1 micron was used to control relative delay of the photons by translating the fiberport couplers along the axis of the beam. 
Initial alignment was performed using a $60\,\mathrm{nm}$ bandpass interference filter to maximize count rates, after which the system was realigned while progressively reducing the spectral bandwidth by exchanging the interference filters as required.  Using this procedure, we measured a biphoton coherence length of approximately $60\,\mu\mathrm{m}$ with a $10\,\mathrm{nm}$ bandpass filter and $20\,\mu\mathrm{m}$ with a $30\,\mathrm{nm}$ bandpass filter, both centered at $810\,\mathrm{nm}$. Fig.~\ref{HOM3mm}  employs two $3\,\mathrm{mm}$-thick BBO crystals, whereas all subsequent measurements were performed using two $0.5\,\mathrm{mm}$-thick BBO crystals operated in a Type I SPDC configuration. We note that using thicker nonlinear crystals results in increased entangled pairs but also increased temporal walk-off and reduced spectral indistinguishability. 

When a sample is introduced into arm $a$ it invariably introduces a delay that will result in a shifted HOM minima.   This was generally compensated by introducing into arm $b$ a dielectric compensator that was either fused silica (for the TGG) or a HWP (for the HWP measurement).  Its length and composition were chosen to roughly match the optical thickness of the sample placed in arm $a$.

\bibliography{main}

\section{Acknowledgments}

This project was first conceived  at the Aspen Center for Physics, which is supported by National Science Foundation grant PHY-2210452.  At JHU, this work is funded by a Brown Investigator Award, a program of the Brown Institute for Basic Sciences at the California Institute of Technology.  We gratefully acknowledge K. Katsumi, C. Overstreet, T. Pittman, and G. Reid for helpful conversations, instrumentation help, and/or careful reading of this manuscript.

\section{Author Contributions}

N.P.A. conceived this project.   The spectrometer was designed and constructed by V.V.D.  Data was taken and analyzed by V.V.D.   Both authors wrote the manuscript.

\end{document}